\pdfoutput=1
\documentclass[a4paper,12pt]{spieman}  
\usepackage{amsmath,amsfonts,amssymb}
\usepackage{graphicx}

\title{Point spread function reconstruction for SOUL+LUCI LBT data}

\author[a,*]{Matteo Simioni}
\author[a]{Carmelo Arcidiacono}
\author[b]{Roland Wagner}
\author[a]{Andrea Grazian}
\author[a]{Marco Gullieuszik}
\author[c]{Elisa Portaluri}
\author[a]{Benedetta Vulcani}
\author[a]{Anita Zanella}
\author[d]{Guido Agapito}
\author[e]{Richard Davies}
\author[f]{Tapio Helin}
\author[g]{Fernando Pedichini}
\author[g]{Roberto Piazzesi}
\author[d]{Enrico Pinna}
\author[b,h]{Ronny Ramlau}
\author[d]{Fabio Rossi}
\author[f]{Aleksi Salo}
\affil[a]{INAF - Osservatorio Astronomico di Padova, Vicolo dell'Osservatorio 5, Padova, Italy, I-35122} 
\affil[b]{Industrial Mathematics Institute, Johannes Kepler University Linz, Altenberger Strasse 69, Linz, Austria, 4040} 
\affil[c]{INAF - Osservatorio Astronomico d'Abruzzo, Via Mentore Maggini, Teramo, Italy, I-64100} 
\affil[d]{INAF - Osservatorio Astrofisico di Arcetri, Via E. Fermi 5, Firenze, Italy, I-50125} 
\affil[e]{MPE - Max-Planck-Institut für extraterrestrische Physik Giessenbachstrasse 1, Garching, Germany, D-85748} 
\affil[f]{LUT University, P.O.Box 20, Lappeenranta, Finland, FI-53851} 
\affil[g]{INAF - Osservatorio Astronomico di Roma, Via Frascati 33, Monte Porzio Catone, Italy, I-00078} 
\affil[h]{RICAM - Johann Radon Institute for Computational and Applied Mathematics, Altenberger Strasse 69, Linz, Austria, 4040} 

\begin{document} 
\maketitle

\begin{abstract}
This paper presents the status of an ongoing project aimed at developing a PSF reconstruction software for adaptive optics (AO) observations. In particular, we test for the first time the implementation of pyramid wave-front sensor data on our algorithms. As a first step in assessing its reliability, we applied the software to bright, on-axis, point-like sources using two independent sets of observations, acquired with the single-conjugated AO upgrade for the Large Binocular Telescope.

Using only telemetry data, we reconstructed the PSF by carefully calibrating the instrument response. The accuracy of the results has been first evaluated using the classical metric: specifically, the reconstructed PSFs differ from the observed ones by less than $2\%$ in Strehl ratio and $4.5\%$ in full-width at half maximum. Moreover, the recovered encircled energy associated with the PSF core is accurate at $4\%$ level in the worst case. The accuracy of the reconstructed PSFs has then been evaluated by considering an idealized scientific test-case consisting in the measurements of the morphological parameters of a compact galaxy.

In the future, our project will include the analysis of anisoplanatism, low SNR regimes, and the application to multi-conjugated AO observations.
\end{abstract}


{\noindent \footnotesize\textbf{*}Matteo Simioni,  \linkable{matteo.simioni@inaf.it}}

\section{Introduction}\label{sec:intro}

Next generation workhorse facilities for 8-10m class ground-based telescopes [e.g., ERIS (Enhanced Resolution Imager and Spectrograph) and MAVIS (MCAO-Assisted Visible Imager and Spectrograph) at the Very Large Telescope; SHARK-visible (System for coronagraphy with High order Adaptive optics from R to K band) and SHARK-near-infrared (NIR) at the Large Binocular Telescope (LBT); KAPA (Keck All-sky Precision Adaptive-optics) and KPF (Keck Planet Finder) at the W. M. Keck Observatory] will improve Adaptive Optics (AO) assisted imaging and spectroscopy, both in the visible and at NIR wavelengths. The great advances obtained by these AO systems will pave the road for extremely large telescopes (e.g., the Giant Magellan Telescope, the European Southern Observatory’s Extremely Large Telescope (ELT) and the Thirty Meter Telescope) in the future to obtain sharper and deeper images than the ones obtained with big space telescopes (e.g., Hubble Space Telescope and James Webb Space Telescope), exploiting all their ground-breaking imaging capabilities.

In many scientific applications exploiting diffraction limit AO capabilities\cite{2000PASP..112..315W, 2012ARA&A..50..305D, 2016SPIE.9908E..1ZD, 2018SPIE10702E..09D, 2020SPIE11448E..2WA, 2020SPIE11447E..1RR} , a serious challenge experienced during data analysis is the precise knowledge of the two-dimensional point spread function (PSF): approaching the diffraction limit of ground-based cornerstone facilities is not an easy task. A severe drawback in this sense is that the PSF of an AO system is a very complex entity, which has a high number of degrees of freedom involved, varies spatially in the field of view, is not constant in time, and behaves differently at each wavelength. Accurate knowledge of the PSF is mandatory to perform high-quality photometric, spectroscopic, and astrometric analysis of astronomical AO data-sets. For instance, a detailed knowledge of the PSF is an unavoidable ingredient to carry out transformational science cases e.g. resolved stellar clusters beyond the Milky Way, stellar populations beyond the Local Group up to the Virgo cluster, galaxy size evolution, stellar mass growth across cosmic time, substructures in intermediate-$z$ galaxies, detailed studies of high-redshift galaxies, and the formation of the smallest galaxies and proto-globular clusters at high $z$. In addition, a precise reconstruction of the PSF for astronomical images is fundamental e.g. for: i) quality control of scientific frames (especially blank extra-galactic fields); ii) flux calibration of spectroscopic data (since the PSF depends on the wavelength); iii) unbiased photometry in crowded fields; iv) image restoration (e.g. Richardson-Lucy deconvolution); v) forward modelling.  
Currently there is  a growing interest in the development of tools to infer the observed PSF from AO systems. The three main approaches of PSF estimation are 1) based on focal-plane data (e.g. PSF classification, PSF fitting/interpolation, PSF marginal estimation and deconvolution, empirical PSF derivation); 2) based on pupil-plane data [e.g. PSF simulation, PSF reconstruction (PSF-R)]; 3) hybrid techniques that combine the use of point sources in the observed scientific (co-added) images, models of atmospheric turbulence, and instrumental aberrations, hence optimizing the reconstruction (e.g. adaptive PSF-R, myopic deconvolution, calibrated PSF-R). All these methods have been extensively described in Ref.~\citenum{2020SPIE11448E..0AB}.
Focal-plane based methods by default include the non common path aberrations (NCPA) components, which means that imperfections in the optics of the instruments are automatically included in the reconstruction. But they are also sensitive to the image noise, and they can be considered representative of a specific positions in the field and/or for a particular wavelength. Moreover, they require the presence of at least one non-saturated point-like source in the imaged field of view. Pupil-plane based methods, instead, refer to techniques aimed at deriving the post-AO PSF only from telemetry control loop data, e.g. time series of wavefront sensor (WFS) measurements or applied deformable mirror commands and associated calibration, without any recourse to focal plane data. NCPA or segment differential piston should be taken into account separately, following dedicated independent calibrations. Testing the result of pupil-plane methods can only be carried out via comparison of PSFs in the science data.

PSF-R is a versatile and flexible approach, that can also be the initial step for hybrid or adaptive methods.
The PSF-R technique is the only viable possibility when no other solutions are possible, due to several reasons: e.g. there are no suitable point sources in the science frames, the target is faint and complex, as in dense and crowded fields (i.e. the core of globular clusters, where only the PSF cores are detected for the majority of the stars), the bright stars are saturated, the integration times are relatively short, observations have been carried out in poor atmospheric conditions or faint guide stars have been used.
Indeed, for both single conjugated adaptive optics  (SCAO\cite{babcock53}) and multi conjugated adaptive optics (MCAO\cite{becker88,2018ARA&A..56..277R}) observing modes, the AO reference star can be out of the instrumental field of view. Moreover, in laser assisted \cite{foy85} MCAO mode, only faint tip-tilt stars are usually present. 
PSF-R is particularly useful in spectroscopic mode, where no PSF is available in the focal plane, or in non-sidereal tracking mode, where the stars in the field of view may be blurred along one axis and so cannot be used as a PSF reference, or where the Solar System targets (used also as AO reference) can be spatially extended.

The PSF-R approach is particularly convenient for the extra-galactic targets where the field of view is typically void of bright point sources, necessary to characterize the PSF of the science data using other methods. As an example, simulations of stellar densities with TRILEGAL \cite{2005A&A...436..895G} code indicate that, in a typical extra-galactic field, one can expect on average $\sim3$ stars per square arcmin in the range Ks=16-25 Vega mag. 
Considering the typical field of view of AO assisted instruments (of the order of $1\,{\rm arcmin}^2$), the statistical probability of having at least one available star is less than 35\% to 45\%. The PSF-R remains the unique way forward.

In this paper, we present the first results of our ongoing project aimed at obtaining a PSF-R tool, i.e. the estimation of the PSF using only AO control-loop telemetry, following the seminal work of Ref.~\citenum{1997JOSAA..14.3057V}.  
The adopted PSF-R method is described in Ref.~\citenum{2018JATIS...4d9003W} and involves several approximations to reconstruct the PSF from the residual phase power spectrum density. Such algorithm was previously tested on simulated data only. Therefore, the present work is the first application of the method to real telescope data and is meant to validate the approach. For this reason, we limited the present analysis to high signal-to-noise ratio (SNR), on-axis observations to reduce the degree of freedom of the problem and avoid degeneracies in multi-parameter space (e.g. when taking into account wider AO reference SNR range, off-axis angles, seeing values).
We investigate the performance of the selected software with archival, SCAO data from the Large Binocular Telescope (LBT\cite{lbt}). Specifically, the data consist of two distinct set of observations of bright, on-axis point sources, preliminary described in Ref.~\citenum{2020SPIE11448E..37S}. They both have been taken as part of the Single conjugated adaptive Optics Upgrade for LBT (SOUL\cite{2016SPIE.9909E..3VP}) validation operations, using LBT Utility Camera in the Infrared (LUCI\cite{2003SPIE.4841..962S}). 
Moreover, to quantify and compare the level of accuracy obtained, we present an idealized scientific case in the area of applicability of the PSF-R, showing the gain offered by this technique to the analysis of the photometry and morphology of a high-$z$ galaxy.

The paper is organized as follows: 
the PSF-R algorithm is described in Section \ref{sec:psfr}; the data are presented in Section \ref{sec:data}; in Section \ref{sec:res} the performance of the PSF-R algorithm is discussed evaluating also the impact on the measure of the morphological parameters of compact galaxies, and Section \ref{sec:fin} is dedicated to the summary and conclusions.
Finally, an interesting relation linking the encircled energy (EE), the full width at half maximum (FWHM), and Strehl ratio (SR) of the observed PSF is discussed in Appendix \ref{sec:sr_vcorr}.


\section{PSF-R method}\label{sec:psfr}

In this section, we describe the method used to reconstruct the on-axis PSF from AO telemetry data, the one described in Ref.~\citenum{2018JATIS...4d9003W}, which is by itself an update of the classical algorithm developed by Ref.~\citenum{1997JOSAA..14.3057V}.
As detailed in Ref.~\citenum{2018JATIS...4d9003W}, the algorithm is highly flexible with respect to the WFS, and we performed minor tuning to adapt it to SOUL+LUCI. Specifically, to our knowledge this is the first successful implementation of a PSF-R algorithm to a pyramid WFS AO system. This is especially relevant also in sight of the upcoming next generation AO instruments that will make wide use of this kind of WFS.
The core idea of this approach is to compute the Optical Transfer Function (OTF), the Fourier transform of the PSF derived from the pupil mask. This is advantageous because the OTF can subsequently be factorized into several independent components (dependent on wavelength and line-of-sight), which can be either calculated from the wavefront residual phase or using tabulated values derived from appropriate calibrations.

The OTF can be defined as:
\begin{equation}\label{eq:otf_tot}
OTF_{AO+NCPA} = OTF_{AO}\cdot K_{NCPA}
\end{equation}
where $OTF_{AO}$ is the post-AO OTF  describing the effects of the AO corrected atmospheric turbulence and $K_{NCPA}$ is the non-AO filter  describing the effects due to the imperfection of the optical system. 

Following the mathematical analysis of Refs.~\citenum{1997JOSAA..14.3057V}~and~\citenum{2018JATIS...4d9003W}, $OTF_{AO}$ is described as: 
\begin{equation}
OTF_{AO} = OTF_{tel} \cdot \exp (-\frac{1}{2}D)
\end{equation}
where $D$ is the structure function of the residual incoming phase (i.e. a  time average of the spatial correlation\cite{2018JATIS...4d9003W})  and thus wavelength dependent, while $OTF_{tel}$ is the OTF of the telescope computed from the pupil mask. $D$ is split into two components that are independent as they describe the part corrected by the AO system, $D_\|$ (lower order structure function), and the part perpendicular to the AO corrected one, $D_\perp$ (higher order structure function). 
From these two structure functions, the associated filters $K_{\|}$ and $K_{\perp}$ can be computed as
\begin{equation}\label{eq:sffilt}
K_x = \exp (-\frac{1}{2} D_x)
\end{equation}
with $x \in \{\|, \perp\}$.
This gives:
\begin{equation}\label{psfr:OTF_mult}
     OTF_{AO} = OTF_{tel} \cdot K_{\|} \cdot K_{\perp}.
\end{equation}
Therefore, if we are able to measure the $D$ components from telemetry data, we can go back to the structure function and  determine the OTF (and consequently the PSF) of the system.

To compute $D_\|$, one needs the residual phase in the space of the mirror modes. However, we only have an estimate for it, namely the reconstructed residual phase computed from the WFS data, which is perturbed by noise and aliasing effects. Therefore, since the assumptions from Refs.~\citenum{1997JOSAA..14.3057V}~and~\citenum{2018JATIS...4d9003W} of fast frame rate and a least squares reconstructor are fulfilled, we can use the same decomposition as in Ref.~\citenum{1997JOSAA..14.3057V} to decompose $D_\|$ further as
\begin{equation}\label{eq:dpar}
D_\| = D_{REC} - D_{n} + D_{al}
\end{equation}
where $D_{REC}$ is the structure function computed from the reconstructed residual incoming phase $\phi_{REC,t}$, $D_n$ is the noise structure function and $D_{al}$ is the aliasing structure function. It is worth noting that this decomposition is valid independently of the type of WFS. As detailed in Ref.~\citenum{2018JATIS...4d9003W}, the adopted method takes as input the reconstructed residual phases $\phi_{REC,t}$, which are recovered from the applied updates of the deformable mirror, to compute $D_{REC}$.
\begin{table}[h]
\centering
\caption{Structure functions computed for the PSF-R. A description on how they are obtained is also provided.}
\label{tab:strfunc}
\begin{tabular}{p{.07\columnwidth}p{.2\columnwidth}p{.58\columnwidth}}
\hline\hline
            & Name                                     & Description \\
\hline
$D_{\perp}$ & Wavefront higher order structure function & Computed using the corresponding seeing $r_0$ and pupil mask via a statistical model of the atmospheric turbulence (von Karman, Kolmogorov). \\[0.5em]
$D_{REC}$   & Residual structure function  & Computed from the reconstructed residual incoming phase $\phi_{REC,t}$, for time frames $t$ and related to the direction of the NGS at wavelength $\lambda$. 
The residual incoming phase is in turn obtained directly from the telemetry. For the pyramid WFS, it has been computed as $\phi_{REC,t}=s\, C\, g\, M$ where $s$ is the WFS measurement, the so-called slopes of the pyramid WFS; $C$ is the control matrix; $g$ is the modal control loop gain vector of the pyramid WFS and $M$ is the description of the mirror modes in terms of phase (a product of the instrument calibtrations). \\[0.5em]
$D_{n}$     & Noise structure function                  & Computed from the noise model. The control matrix, which already incorporate WFS-specific effects, has been used to propagate from WFS level to phase level. \\[0.5em]
$D_{al}$    & Aliasing structure function               & A model is used to propagate simulated higher order parts of the phases through the WFS model and then compute the response of the AO control algorithm for the corresponding seeing $r_0$, the used AO control algorithm, and pupil mask. To retrieve a statistical average, this procedure is rerun for many simulated phases before all the computed aliasing phases are used in the structure function computation.
\\[0.5em]
$D_\|$      & Wavefront lower order structure function & Computed from $D_{REC}$, $D_{n}$, and $D_{al}$ following Eq. \ref{eq:dpar}. \\
\hline
\end{tabular}
\end{table}
Any WFS-specific issues have to be solved already when setting up the AO system control algorithm. In the case of SOUL the optical gain is measured through a low-amplitude sinusoidal modulation of a controlled mirror mode. Thus, the SOUL slopes telemetry already include the correction for the optical gain. Therefore, the used PSF-R algorithm is flexible enough to account for specific characteristics of the pyramid WFS without any major modification.
We refer the interested reader to a more detailed description of the various structure functions provided in Table \ref{tab:strfunc}. 

The last term needed for the computation of the post-AO $OTF$ ($OTF_{AO}$) is $OTF_{tel}$, which is computed from the calibrated pupil mask and then multiplied by the aforementioned filters following Equation \eqref{psfr:OTF_mult}.

$K_{NCPA}$ in Eq. \eqref{eq:otf_tot} is computed from the calibrated phase map and depends on the specific instrument configuration and telescope pointing. It therefore varies in the field.  

The final PSF is then retrieved by transforming the ${OTF}_{AO+NCPA}$  via inverse Fourier transform into the ${PSF}_{REC}$, which is also integrated in wavelength over the pass-band of the selected filter.

We note that focusing only on on-axis point source, the anisoplanatic component can be neglected, thus reducing the degrees of freedom in the PSF-R. Moreover, considering that during the observations described in this work the loop was sampled at frequencies $>500$ Hz, and we expect that non-negligible physical vibrations occur at much slower variations, we can assume that the SOUL+LUCI system already accounted for all these variations. Therefore we do not need to account for any additional term related to vibrations, as it is already present in the used data.


\section{Data}\label{sec:data}

As the main goal of this work is to apply the selected algorithm for PSF-R to real AO data, we used SCAO near infrared images and associated synchronous telemetry acquired with the SOUL system mounted on the LBT and feeding the LUCI near-infrared camera. This, in turn, is equipped with three optical zoom configurations delivering the same number of distinct resolutions (see Ref.~\citenum{2021MNRAS.508.1745A} for a detailed description of the instrument set-up). We use the diffraction limited mode, called \textrm{N30}, a $2048\times2048\,{\rm px}$ configuration with $30"\times 30"$ field of view ($14.95\, {\rm mas/px}$). 

\begin{table}[htp]
\caption{Log of the SOUL+LUCI@LBT observations} 
\label{tab:log}
\centering    
\begin{tabular}{lcc}
\hline\hline
                         & DAYTIME             & NIGHTTIME          \\
\hline
PROG. ID                 & 1183150             & 1286078            \\
Target                   & illuminated fiber   & HD$\,873$          \\
RA [hh mm ss]            & N.A.                & $00\, 13\, 13.123$ \\
DEC [$^\circ \, ' \, ''$] & N.A.               & $+20\, 45\, 9.479$ \\
R [mag]                  & $10.0$              & $8.57$             \\
Date [yyyy/mm/dd]        & 2019/03/29          & 2019/11/09         \\
Filter                   & H                   & FeII               \\
NDIT$\times$DIT [s]      & $90\times1$         & $20\times0.313$    \\
AIRMASS                  & N.A.                & $1.0248$           \\
\hline
\end{tabular}
\end{table}

We used archive data, consisting of 2 independent datasets. The first set refers to observations of an artificial source (i.e. an illuminated fiber - hereafter "daytime" dataset) with controlled atmospheric condition \cite{2008SPIE.7015E..12R} . The second dataset consists of observations of a natural point source (i.e. a bright star - hereafter "nighttime" dataset). The log of the observations is presented in Table \ref{tab:log}.

\subsection{Observed PSFs}
The LUCI data reduction procedure is identical for both datasets. 
First, a bad-pixel map is obtained from the dark frames and a background image is generated averaging the sky frames. 
Individual science frames have been stacked with a sigma-clipped mean to increase the final SNR; we have then subtracted the background image from the stacked science one.
Finally, we masked the bad pixels and estimated their values by linear interpolation of the neighboring ones.
We corrected the obtained images for the residual background and cropped them, selecting a region of $\sim8\,{\rm arcsec^{2}}$ ($190{\rm px}\times190{\rm px}$), centered at the point source. Finally, we normalized them to their total flux.
These high SNR images will be referred in the following as  ``observed PSFs''.

\subsection{Reconstructed PSFs}

In the following, we provide additional technical details on the application of the adopted PSF-R method in the specific case of SOUL+LUCI data.
We only used the telemetry data associated with the scientific observations to reconstruct the PSFs.

The telemetry data provided by the SOUL team include: AO WFS slopes data history, control matrix, interaction matrix, gain vector, and the pupil definition. In particular, we obtained pupil images taken with LUCI in the "pupil image" mode used for the alignment of the instrument. This imaging mode unfortunately does not include the additional pupil obstruction exclusive of the N30 camera. 
We derived them from the camera mechanical drawing, to obtain the pupil shape for daytime dataset.
For nighttime, additional contributions have been taken into account in defining the pupil shape. In this configuration, in fact, LUCI sees also the secondary mirror mounting (including the LBT swing arm) shadow occurring in the pupil mask. While the N30 camera system is fixed with the LUCI detector, the swing arm projection depends on the orientation of the LUCI camera that is continuously updating to follow the apparent rotation of the sky. The derived pupil shape is used as a binary mask assuming negligible modulation of the amplitude (optical system transmission) over the aperture. Using the LUCI images we verify the pixelscale value which results $14.95\,{\rm mas/px}$ for both the H and FeII filters.

\begin{figure}
  \centering
     \includegraphics[width=0.9\columnwidth]{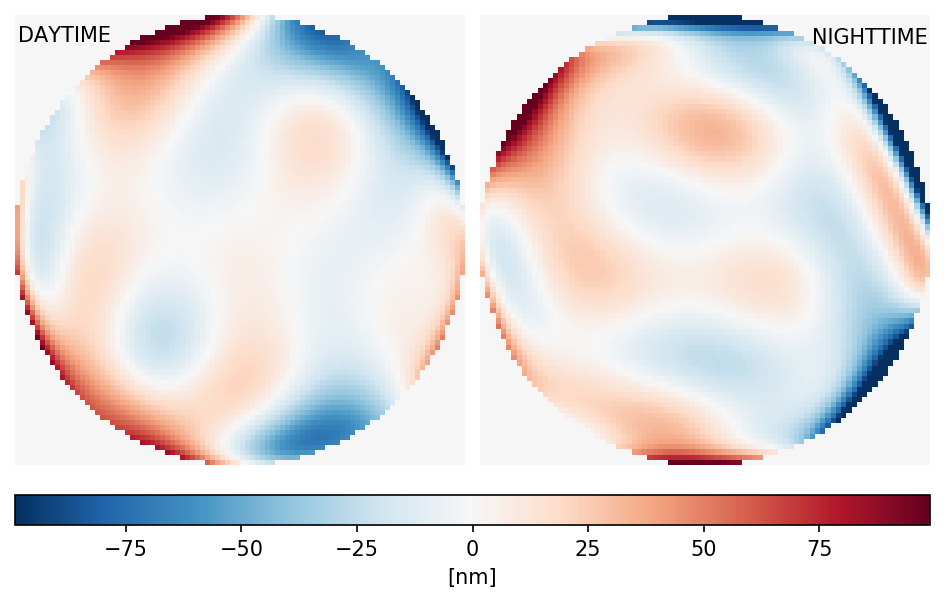}
     \caption{NCPA mean residuals, after AO correction, for daytime (left) and nighttime (right). The pixel values are in ${\rm nm}$.}
     \label{fig:pmncpa}
\end{figure}

In all the considered observations, NCPA between WFS optical train and the LUCI camera are compensated by the AO system making the pyramid loop closing on non-zero reference slopes. This correction, however, leaves residual depending on the pyramid optical gain \cite{Korkiakoski} and un-calibrated reciprocal flexures of the WFS with respect to LUCI. These residuals may count up to $100\,{\rm nm}$ \cite{2021MNRAS.508.1745A} of wavefront error. Lacking proper calibrations, we  measured these residuals by analyzing the observed PSF: this method, applied on long exposure PSF, has low sensitivity. In the present case we observed residuals (tip-tilt free) accounting for about $40\,{\rm nm}$ in both daytime and nighttime. The measured phase maps for NCPA are shown in Figure \ref{fig:pmncpa}. As a consequence, the small residual values 
also indicate that the pyramid WFS gain measures registered in the telemetry data are accurate. Even if negligible, these estimated wavefront errors were also applied in the PSF-R algorithm.

\subsection{Diffraction limited PSFs}

We numerically computed the monochromatic diffraction limited PSF thanks to the knowledge of the pupil shape. This was obtained as the square modulus of the Fourier transform of the pupil mask, which was properly zero-padded to obtain the adequate pixelscale. Finally, different sizes of the zero-padding were used to take into account the whole passband of the selected LUCI filters. For both datasets, the final diffraction limited PSFs are composed by $11$ monochromatic ones, calculated at discrete wavelength intervals over each passband.

\subsection{Daytime}
\begin{figure*}
  \centering
     \includegraphics[width=0.9\textwidth]{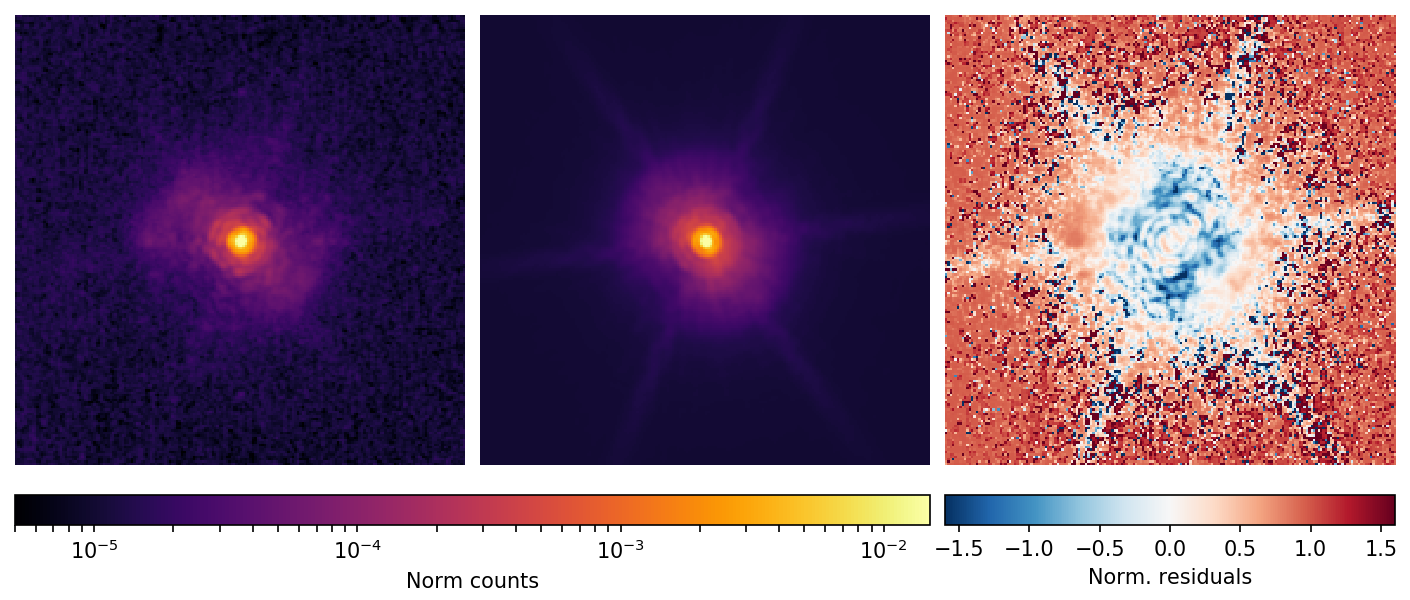}
     \caption{Daytime PSFs. Left panel: observed PSF. Central panel: reconstructed PSF. Both are in arbitrary logarithmic color scales. Right panel: normalized residuals between observed and reconstructed PSF. The residuals are normalized pixel-per-pixel with the observed PSF. The color scale is linear and the colorbar is on the bottom. There are no evident residual structures of the two PSFs. All the images are $\sim8\,{\rm arcsec^{2}}$ ($190{\rm px}\times190{\rm px}$).}
     \label{fig:dpsf}
\end{figure*}

For this particular dataset, laboratory-like conditions have been used: optical turbulence disturbance has been generated and corrected \cite{2010SPIE.7736E..09E} using the Adaptive Secondary \cite{2003SPIE.5169..159R} of the LBT, with the telescope and AO system configured for daytime \cite{2008SPIE.7015E..12R} . Specifically, a seeing of $1.2$ arcsec was simulated with an average wind speed of $15\,{\rm m/s}$ \cite{2021MNRAS.508.1745A} . The simulated point source has an equivalent R-band magnitude of $10.0\,{\rm mag}$.
Three sets of $30$ $1$s-images have been obtained in H band ($1.5-1.8\,{\rm \mu m}$, centered at $1.65\,{\rm \mu m}$), achieving a temporal frequency for the AO correction of $500\,{\rm Hz}$. The resulting saving rate for the associated telemetry is $500\,{\rm frames/s}$. In all the science frames the point source is not saturated, within the linear regime of the detector. The resulting SNR per pixel for the observed PSF goes from $250$ at peak to a mean value of $10$ at a radial distance of $300$ mas ($20{\rm px}$) from PSF centre and reaching mean SNR per pixel$=1$ around $0.9$ arcsec ($60$px).
The observed PSF for daytime is presented in the left panel of Figure~\ref{fig:dpsf}. The measured SR - defined as the ratio between the peak values of a selected PSF and of the diffraction limited one - is $0.59$ and the FWHM is $2.95\,{\rm px}$. The central panel of Figure~\ref{fig:dpsf} shows the reconstructed PSF. The match between the two PSFs can be qualitatively inferred from the right panel of the same figure, where the normalized residuals are presented. No evident residual PSF structures are present.
A quantitative assessment on the performance of the reconstructed PSF is presented in the following sections.

\subsection{Nighttime}
\begin{figure*}
  \centering
     \includegraphics[width=0.9\textwidth]{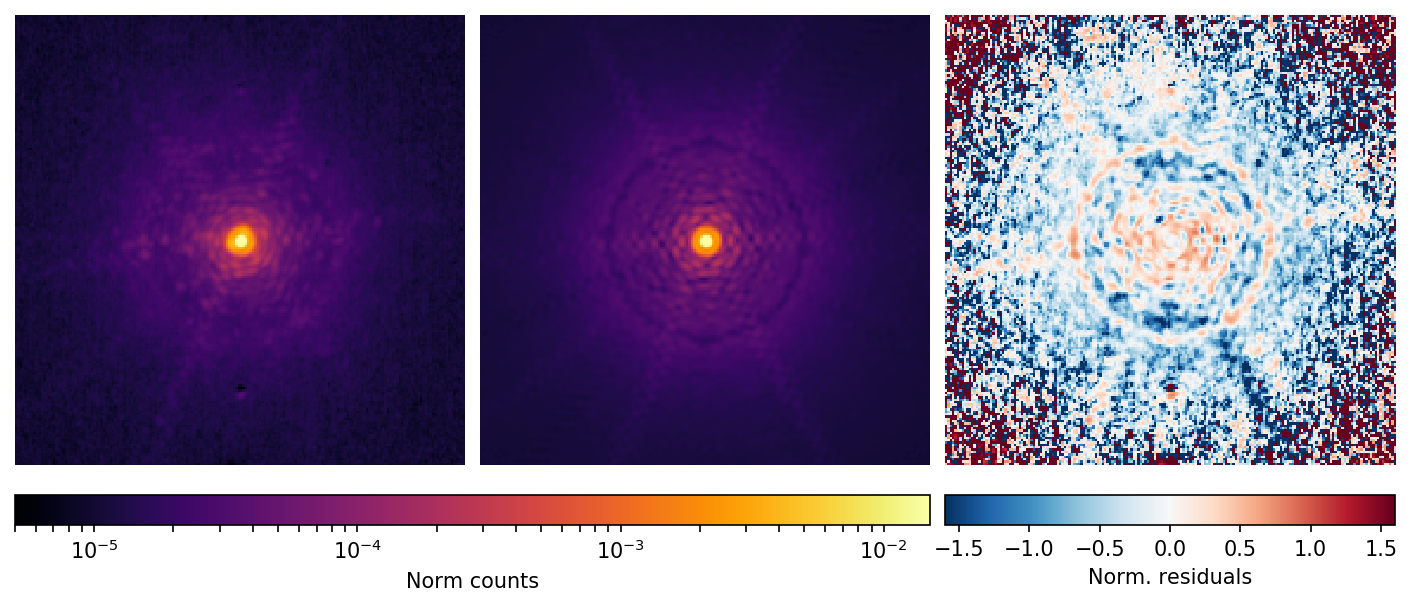}
     \caption{Nighttime PSFs. Left panel. Observed PSF. Central panel. Reconstructed PSF. Right panel. Normalized residuals between observed and reconstructed PSF. Images size, color codes and scales are the same as in Figure \ref{fig:dpsf}.}
     \label{fig:npsf}
\end{figure*}

The nighttime dataset has been obtained in the narrow-band FeII filter, centered at $1.65\,{\rm \mu m}$, with FWHM $0.02\,{\rm \mu m}$. The use of a narrow-band filter reduces the impact of atmospheric dispersion on data: LUCI is not equipped with an atmospheric dispersion corrector and the associated blur of the PSF is of the order of few pixels, so we properly take into account the effect while reconstructing the PSF. The dataset consists of on-axis observation of a bright star (HD$\,873$, R=$8.57\,{\rm mag}$). The star has been also used as AO reference, thus obtaining the best correction from the AO system along the direction of the target. In this case, $20$ images of $0.313\,{\rm s}$ have been considered. The measured seeing was $1.20$ arcsec and the wind speed $3.4\,{\rm m/s}$. Using a bright target, the frequency of AO correction could be raised to $1700\,{\rm Hz}$, with a resulting saving rate for the associated telemetry of $850\,{\rm frames/s}$ (decimated by a factor 2). The measured SR is $0.58$ and the FWHM $43.95\,{\rm mas}$ ($2.93\,{\rm px}$).  Also in this case, all the images of the point source are not saturated and within the linear regime of the detector. The final SNR per pixel of the observed PSF is $50$ at the peak, decreasing to a mean value of $10$ at $300$ mas. For this dataset, a mean SNR per pixel$=1$ is reached around $1.3$ arcsec ($90$px) from PSF centre. The difference between the SNR radial trend of the two dataset is due to the different bandwidth of the filters used for each dataset: a broadband one for daytime and a narrowband one for nighttime. The variation in environmental conditions during the observations is also a contributing factor.
The observed PSF is shown in the left panel of Figure \ref{fig:npsf}, the reconstructed one in the central panel. In the reconstructed PSF it can be noted a feeble ring-like structure, with a radius of $\sim0.6\,{\rm arcsec}$ ($\sim40\,{\rm px}$). This approximately coincides with the expected location of the control radius, which delimits the region of the PSF affected by AO correction. This ring-like structure leaves a trace also in the residuals, shown in the right panel of Figure \ref{fig:npsf}, and it is due to the method used in the adopted PSF-R algorithm to treat corrected spatial modes.


\section{Results and Discussion}\label{sec:res}
\begin{table}[ht]
\centering 
\caption{Properties of the observed and reconstructed PSFs for both datasets. See text for details.}
\label{tab:orprop}
\begin{tabular}{lccc}
\hline\hline
Parameter           & OBS.  & REC.  & DIFF. $[\%]$ \\
\hline
DAYTIME             &       &       &              \\
\hline
SR                  & 0.59  & 0.58  & 1.7          \\
FWHM[mas]           & 44.10 & 43.80 & 0.7          \\
${\rm EE_{core}}$   & 0.43  & 0.43  & 1.9          \\
${\rm R_{50}}$[mas] & 50.23 & 53.97 & 7.4          \\
$\chi^{2}_{RED}$    &       & 4.63  &              \\
Corr. with OBS.     &       & 0.996 &              \\
\hline
NIGHTTIME           &       &       &              \\
\hline
SR                  & 0.58  & 0.59  & 1.7          \\
FWHM[mas]           & 43.80 & 45.75 & 4.4          \\
${\rm EE_{core}}$   & 0.43  & 0.45  & 4.0          \\
${\rm R_{50}}$[mas] & 52.92 & 50.83 & 3.9          \\
$\chi^{2}_{RED}$    &       & 3.27  &              \\
Corr. with OBS.     &       & 0.995 &              \\
\hline
\end{tabular}
\end{table}

In this section, we detail the performance of the adopted PSF-R algorithm.
A qualitative analysis of Figures \ref{fig:dpsf} and \ref{fig:npsf} corroborates the good match between the observed and reconstructed PSF. In particular, the residuals shown on the right panels suggest that the core of the PSF is well reproduced. More quantitatively, a collection of the most relevant properties of the observed and reconstructed PSFs is provided in Table \ref{tab:orprop}. For each dataset and PSF, we report the measured SR and FWHM. The EE - defined as the amount of PSF energy that is contained within a fixed circular aperture, centered on the PSF center - in the core of the PSF, ${\rm EE_{core}}$, is also listed. The radius of the aperture, in this case, has been fixed to $40.4$ mas ($2.7\,{\rm px}$): the expected FWHM of the diffraction limited PSF. ${\rm R_{50}}$ is the aperture (in mas) that contains $50\%$ of the total light. 
\begin{table}[ht]
\centering
\caption{As in Table \ref{tab:orprop} but for the diffraction limited case and the best fit to the observed PSF with an analytic profile that is the sum of 2 Moffat functions.}
\label{tab:dlmc}
\begin{tabular}{p{2.3cm}p{.7cm}p{.7cm}p{.7cm}|p{.7cm}p{.7cm}}
\hline\hline
Parameter           & OBS   & D.L.  & DIFF. [\%] & Moffat & DIFF. $[\%]$ \\
\hline
DAYTIME             &       &       &            &        &              \\
\hline
SR                  & 0.59  & 1.00  &            & 0.61   & 3.4          \\
FWHM[mas]           & 44.10 & 40.36 & 8.5        & 44.10  & 0.0          \\
${\rm EE_{core}}$   & 0.43  & 0.62  & 44.1       & 0.44   & 2.3          \\
${\rm R_{50}}$[mas] & 50.23 & 30.20 & 39.9       & 48.14  & 4.2          \\
$\chi^{2}_{RED}$    &       & 10.70 &            & 4.97   &              \\
Corr. with OBS.     &       & 0.974 &            & 0.997  &              \\
\hline
NIGHTTIME           &       &       &            &        &              \\
\hline
SR                  & 0.58  & 1.00  &            & 0.60   & 3.4          \\
FWHM[mas]           & 43.80 & 40.07 & 8.5        & 43.80  & 0.0          \\
${\rm EE_{core}}$   & 0.43  & 0.62  & 44.1       & 0.44   & 2.3          \\
${\rm R_{50}}$[mas] & 52.92 & 30.50 & 42.4       & 49.78  & 5.9          \\
$\chi^{2}_{RED}$    &       & 8.27  &            & 6.02   &              \\
Corr. with OBS.     &       & 0.982 &            & 0.995  &              \\
\hline
\end{tabular}
\end{table}

Finally, to quantify the match between the observed PSF and the reconstructed one, two parameters are listed in Table \ref{tab:orprop}. The first is $\chi^{2}_{RED}$ which is the reduced $\chi^{2}$ between observed and reconstructed PSFs. It has been obtained using GALFIT \cite{2010AJ....139.2097P} on the observed PSF with the reconstructed PSF as input. The center position of a ``psf'' model has been set as the only free parameter for the fit and an input rms image has been created directly from the science data.
The second parameter is the correlation level between the observed PSF and the reconstructed one. It is measured  pixel-per-pixel over the whole PSF image ($8\,{\rm arcsec^{2}}$) by means of Pearson R index.
For comparison, the same parameter list is also provided in Table \ref{tab:dlmc} for the diffraction limited case. We also include in the table the results associated to the best fit of the observed PSFs with an analytic profile, the sum of 2 Moffat functions. The choice is motivated by the fact that fitting a single Moffat function does not provide a good match on both the core and the wings of the observed PSF simultaneously (Ref.~\citenum{2016SPIE.9909E..5YP} and references therein).

\begin{figure*}
  \centering
     \includegraphics[width=0.9\textwidth]{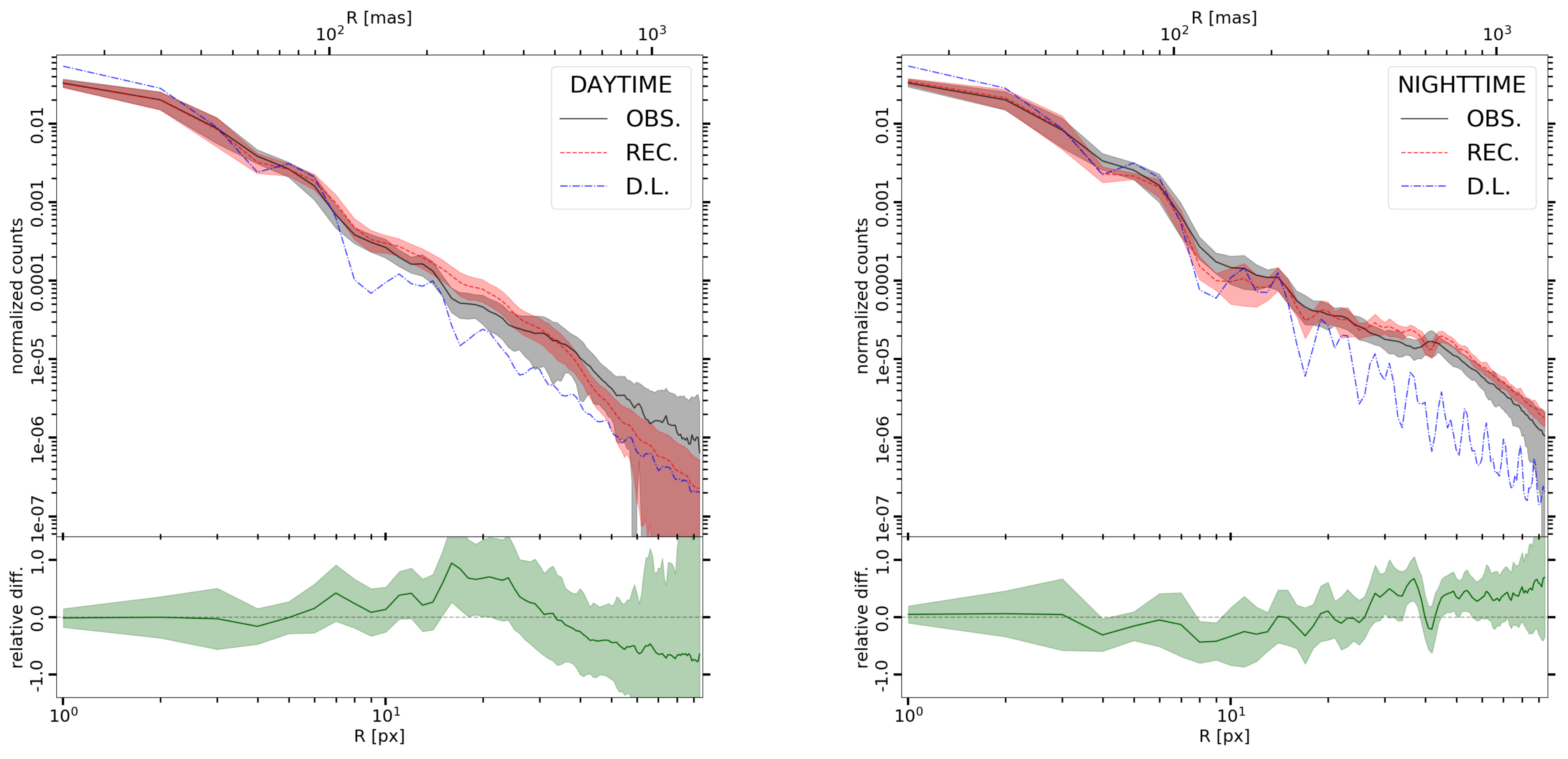}
     \caption{Radial profile (mean over all angles) of the observed PSF (black) and the reconstructed PSF (red). Left panel is for daytime, while the right one is for nighttime. The associated standard deviations are indicated with shaded regions. For comparison, the radial profile of the diffraction limited PSF (D. L.)  is also shown (blue dot-dashed line). The green profile on bottom shows the normalized deviation, as a function of distance, of the reconstructed PSF with respect to the observed one. The associated standard deviation is represented as a shaded area. We computed this quantity as the quadratic sum of the standard deviations associated to observed and reconstructed PSFs.}
     \label{fig:rpdn}
\end{figure*}
Table~\ref{tab:orprop} shows that the reconstructed PSFs match extremely well the observed ones, both for daytime and nighttime data. Indeed, the SR is recovered with an accuracy of $1.7\%$ in both cases. 
To further discuss the other parameters, we compare the radial profile of the observed and reconstructed PSFs. They are reported in Figure~\ref{fig:rpdn} for daytime (left panel) and nighttime (right panel). The radial profile of the diffraction limited PSF is also provided for comparison.
For completeness, the relative differences between observed and reconstructed PSF are also reported on the insets on top.
In both cases, the radial profile of the reconstructed PSF nicely follows the observed one, particularly in the central regions. This is confirmed by the measured FWHM of the reconstructed PSFs that are accurate at the level of $1\%$ and $4\%$ for daytime and nighttime, respectively. This translates into differences of the order of $2\,{\rm mas}$ ($0.13\,{\rm px}$) in the worst case. Even the EE$_{core}$ in the core is fairly reproduced, with a precision better than $4\%$ in both cases. 
Moving outwards to around $50\,{\rm mas}$ ($3.5\,{\rm px}$), the light distribution of the reconstructed PSF keeps following the observed one with errors on ${\rm R_{50}}$ of the order of $7\%$ and $4\%$ for daytime and nighttime respectively ($0.25\,{\rm px}$ in the worst case). For larger apertures, small deviations (relative difference $<1$) are present in the radial profiles of the reconstructed PSFs with respect to the observed ones. In particular, we note the bump in the radial profile of the nighttime reconstructed PSF, associated to the ring-like structure visible in Figure \ref{fig:npsf}. This feature, however, does not seem to affect significantly the match between observed and reconstructed PSF, as demonstrated by the associated low $\chi^{2}_{RED}$ and high correlation level values. Consistent values are found also in the daytime case.

For what concerns the comparison with other PSFs, the diffraction limited PSF does not perform as well as the reconstructed one. All the values reported in Table~\ref{tab:dlmc} indicate larger deviations than what has been obtained with the reconstructed PSFs. Furthermore, in the case of the two-Moffat fit, ${\rm EE_{core}}$ and ${\rm R_{50}}$ are well recovered, as expected, but the $\chi^{2}_{RED}$ and correlation level values indicate that the reconstructed PSFs work equally well or even slightly better than the analytic profile.

\subsection{Impact of the PSF-R on an ideal science case}\label{sec:scieval}

The aim of this work is to obtain a reliable estimate of the differences between the reconstructed PSF and the observed one (as listed in Table \ref{tab:orprop}), and to test the applicability of the reconstruction to an ideal science case. In this section, we aim to define a general method to assess how differences between observed and reconstructed PSF translate into errors on the scientific measurements. In fact, the same information should be used to define constraints on the reconstructed PSFs to meet the requirements for the precision on selected science measurements. For this reason, we limited as much as possible the effects of other factors related to specific observation.
In particular, we envisaged a representative case where the reconstruction of the PSF is critical to reach the scientific goal and take advantage of the knowledge of the observed (true) PSF to assess quantitatively the potential differences that can be revealed as a comparison.

As an example, we consider the case of a distant (e.g. $z \sim 1.5$), bright and compact (e.g. effective radius comparable to the FWHM of the PSF) galaxy. We investigate what accuracy on the structural parameters is achieved when modelling the 2D light profile of the galaxy with GALFIT using our reconstructed PSF model. For simplicity, we consider a galaxy whose light distribution can be described by a S\'ersic profile. We did not simulate galactic substructures that, albeit more realistic, would complicate the interpretation of our results. Similarly, we  chose to simulate a bright galaxy to assess the effects of the faint and outer region of the PSF, which would be hidden with a lower SNR, but are fundamental because these are the most noticeable differences between the observed and the reconstructed PSF.
The parameters of the  simulated  galaxy are: integrated magnitude ${\rm m_{Vega}}=17.5\,{\rm mag}$; S\'ersic index ${\rm n}=3.9$; effective radius $R_e=94\,{\rm mas}$ ($6\,{\rm px}$, corresponding to $\sim2\,{\rm FWHM}$); axis ratio of $0.28$. These values, with the exception of the integrated magnitude, are typical for a red nugget, an early type galaxy, at redshift $z\sim1.5$ \cite{2009ApJ...695..101D} .

To produce the mock observations, we make extensive use of SimCADO \cite{2016SPIE.9911E..24L,2020SPIE11452E..1ZL} , which we tuned accordingly to simulate SOUL+LUCI. 
We considered an exposure time of $54000\,{\rm s}$, subdivided into $180$ frames (DIT) of $300\,{\rm s}$ each.
To ensure a statistical significance of the results, we generated $100$ mock observations of the same model galaxy, adding at each time random white noise.

\begin{figure*}[t]
  \centering
     \includegraphics[width=0.9\textwidth]{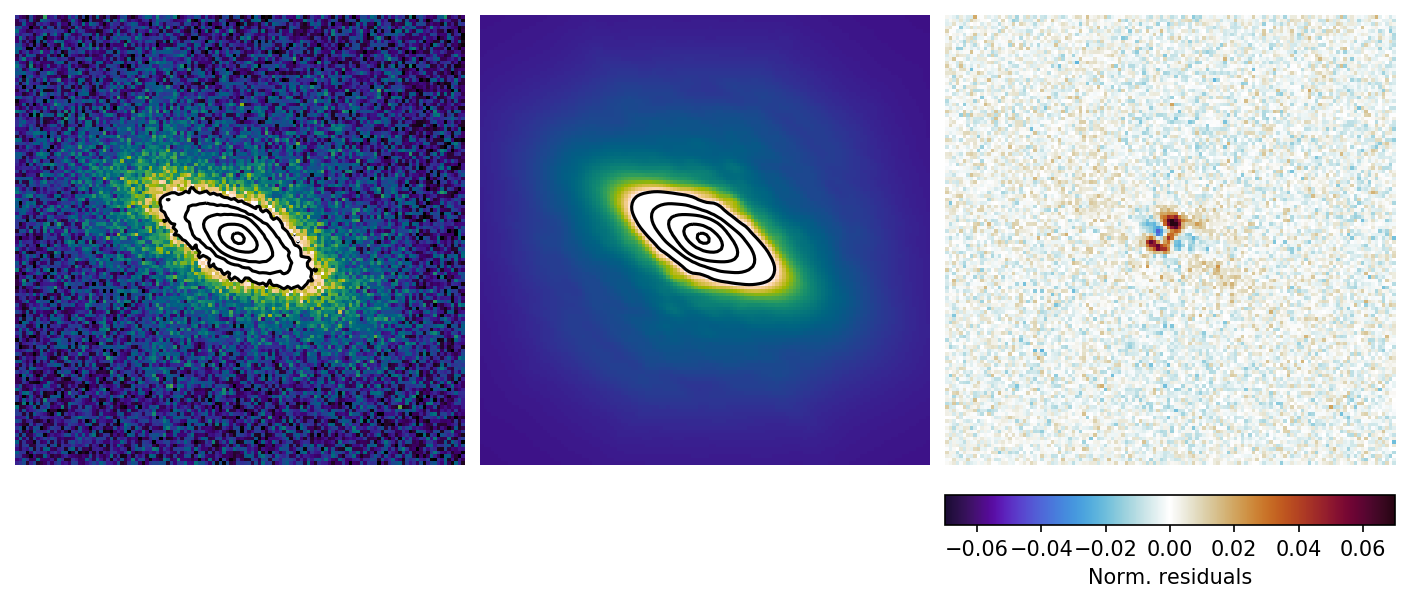} 
     \caption{Left panel. Simulated galaxy. A $3.7\,{\rm arcsec^{2}}$ ($128{\rm px}\times128{\rm px})$ window is shown. The nighttime observed PSF is used as input to generate the simulated observation. Central panel. The best-fit GALFIT S\'ersic model. In this case, the reconstructed PSF is used as the input PSF for GALFIT. Right panel. Normalized residuals. The color scale is linear, the colorbar is shown on the bottom. The normalization has been done dividing the residuals by the simulated observation. It can be noted the excellent agreement between observation and model, despite the two being obtained using the observed and reconstructed PSF as input, respectively. The contours in the left and central panels refer to the same count levels.}
     \label{fig:nscier}
\end{figure*}

The analysis is subdivided in two main parts: the simulation of the  observations and the measurement. The simulated observations have been generated by convolving a pure S\'ersic model with the observed PSF. A zoom-in of the central $128\times128\,{\rm px}$ is presented in the left panel of Figure \ref{fig:nscier}.
The measurements have been performed with GALFIT giving different PSFs as input to the software: i) the observed PSF, for the control measure; ii) the reconstructed PSF, for the actual measure; iii) the diffraction limited PSF, for comparison; iv) no PSF at all, again for comparison. The best-fit GALFIT model (a pure S\'ersic profile) associated to the reconstructed PSF is shown in the central panel of Figure~\ref{fig:nscier}. The contours in the central region of the galaxy mark the same count levels as in the left panel: it can be noted the good agreement between the simulated galaxy and the model. This is emphasized by the normalized residuals, presented in the right panel of Figure~\ref{fig:nscier}. No evident residual structures can be detected.

\begin{table*}[ht]
\centering
\caption{Scientific evaluation results. The mean deviation from input values and associated standard deviation are reported for the main morphological parameters. Specifically, they are: the centre position (x and y), the integrated magnitude, effective radius, S\'ersic index, axis ratio and position angle of the galaxy. The input galaxy is obtained using the observed PSF, while each column refers to the different PSF used by GALFIT to recover the input parameters. 
In particular: the control measure has been obtained using the observed PSF also in the fit, while the one labelled "REC." has been obtained using the reconstructed PSF. For completeness, we reported also the results obtained using the diffraction limited PSF and those without providing any PSF at all as input to GALFIT ("NONE").}
\label{tab:nscier}
\begin{tabular}{lrrrr}
\hline\hline
Parameter         & OBS. [CONTROL]     & REC.               & D.L.               & NONE             \\
\hline
x[mas]             & $ 0.000 \pm 0.164$ & $-0.194 \pm 0.179$ & $-0.164 \pm 0.164$ & $-0.120 \pm 0.284$ \\
y[mas]             & $-0.045 \pm 0.105$ & $-0.030 \pm 0.105$ & $-0.105 \pm 0.105$ & $ 0.000 \pm 0.149$ \\
int. mag. [mag]   & $-0.002 \pm 0.019$ & $-0.009 \pm 0.016$ & $ 0.020 \pm 0.034$ & $ 0.119 \pm 0.017$ \\
${\rm R_{e}}$[mas] & $-0.807 \pm 3.005$ & $ 2.497 \pm 2.272$ & $ 41.845 \pm 11.168$ & $ 69.5773 \pm 6.055$ \\
n                 & $-0.105 \pm 0.324$ & $-0.841 \pm 0.188$ & $ 0.421 \pm 0.486$ & $-1.640 \pm 0.087$ \\
b/a               & $ 0.002 \pm 0.005$ & $ 0.042 \pm 0.005$ & $ 0.103 \pm 0.006$ & $ 0.217 \pm 0.005$ \\
PA[deg]           & $ 0.095 \pm 0.167$ & $-0.621 \pm 0.150$ & $-0.896 \pm 0.232$ & $-0.737 \pm 0.229$ \\
\hline
\end{tabular}
\end{table*}

\begin{figure}[t]
  \centering
     \includegraphics[height=.7\textheight]{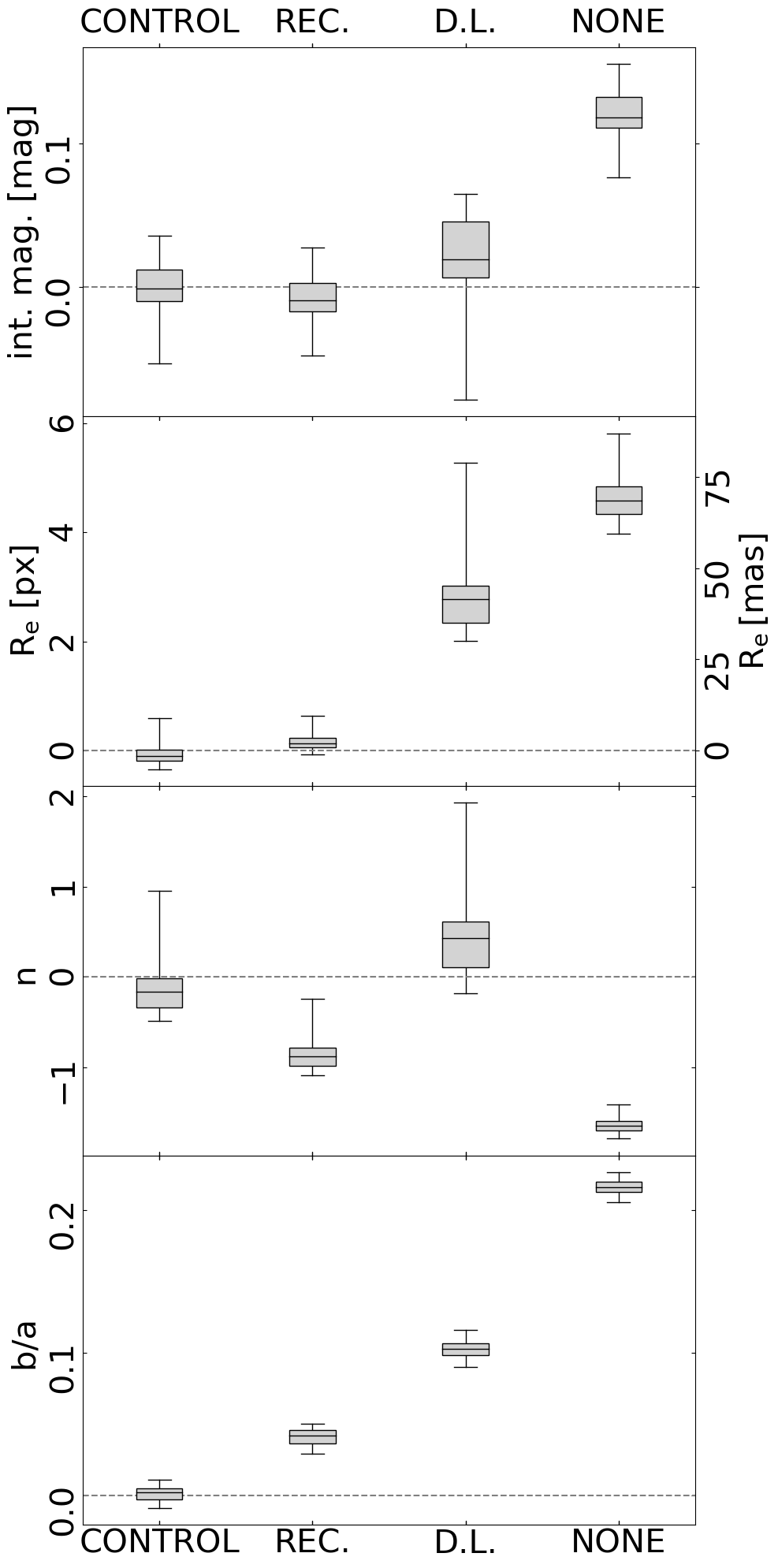} 
     \caption{Box plots of the measured deviations from input values for integrated magnitude, effective radius ${\rm R_{e}}$, S\'ersic index ${\rm n}$, and axis ratio $b/a$ from top to bottom. The control measure has been obtained using the observed PSF (the same used to produce the simulated observations) as input for GALFIT. On the contrary, the actual measure (REC.) has been obtained using the reconstructed PSF as input for GALFIT. For completeness, the box plots are shown also for the case where the diffraction limited PSF is given as input to GALFIT and when no PSF at all was provided as input to the software.}
     \label{fig:sedevbp}
\end{figure}

Quantitatively, all the morphological parameters are in good agreement. We report in Table~\ref{tab:nscier} the mean differences between recovered and input values along with the associated standard deviations. The table includes the results also for the control measure, the diffraction limited case, and the case where no PSF has been provided to GALFIT.
The associated box plots are shown in Figure~\ref{fig:sedevbp}, representing the distribution of the deviations from input values for the $100$ simulations. The main parameters are shown, namely the integrated magnitude, the effective radius ${\rm R_{e}}$, the S\'ersic index ${\rm n}$, and the axis ratio $b/a$ (from top to bottom). 
The loss of accuracy is more noticeable for the case where no PSF at all is provided to GALFIT, confirming the need of the observed PSF in this particular case. 

From Table~\ref{tab:nscier} and Figure~\ref{fig:sedevbp} it can be noted that the recovered S\'ersic index is somewhat biased towards smaller values. The best-fit model obtained providing the reconstructed PSF to GALFIT has a $n$ value $2\sigma$ lower than expected.
As a general consideration, although an optimization of the analysis is possible (but beyond the aim of this study), our two-dimensional fit is performed successfully and the galaxy can be classified in all the cases as an early type object (having $2.5<{\rm n}<4$).

To further assess whether the discrepancy between the measured and intrinsic S\'ersic index is due to our PSF reconstruction, we also applied a mask on both observed and reconstructed PSFs selecting only the pixels inside a given cutting radius (${\rm R_{CUT}}$). We then performed the same analysis described above, using the masked observed PSF to simulate the observations and giving the masked reconstructed one as input to GALFIT to recover the structural parameters of the galaxy. While gradually increasing ${\rm R_{CUT}}$, we record the variations in the measured morphological parameters.

\begin{figure}[t]
  \centering
     \includegraphics[width=.7\columnwidth]{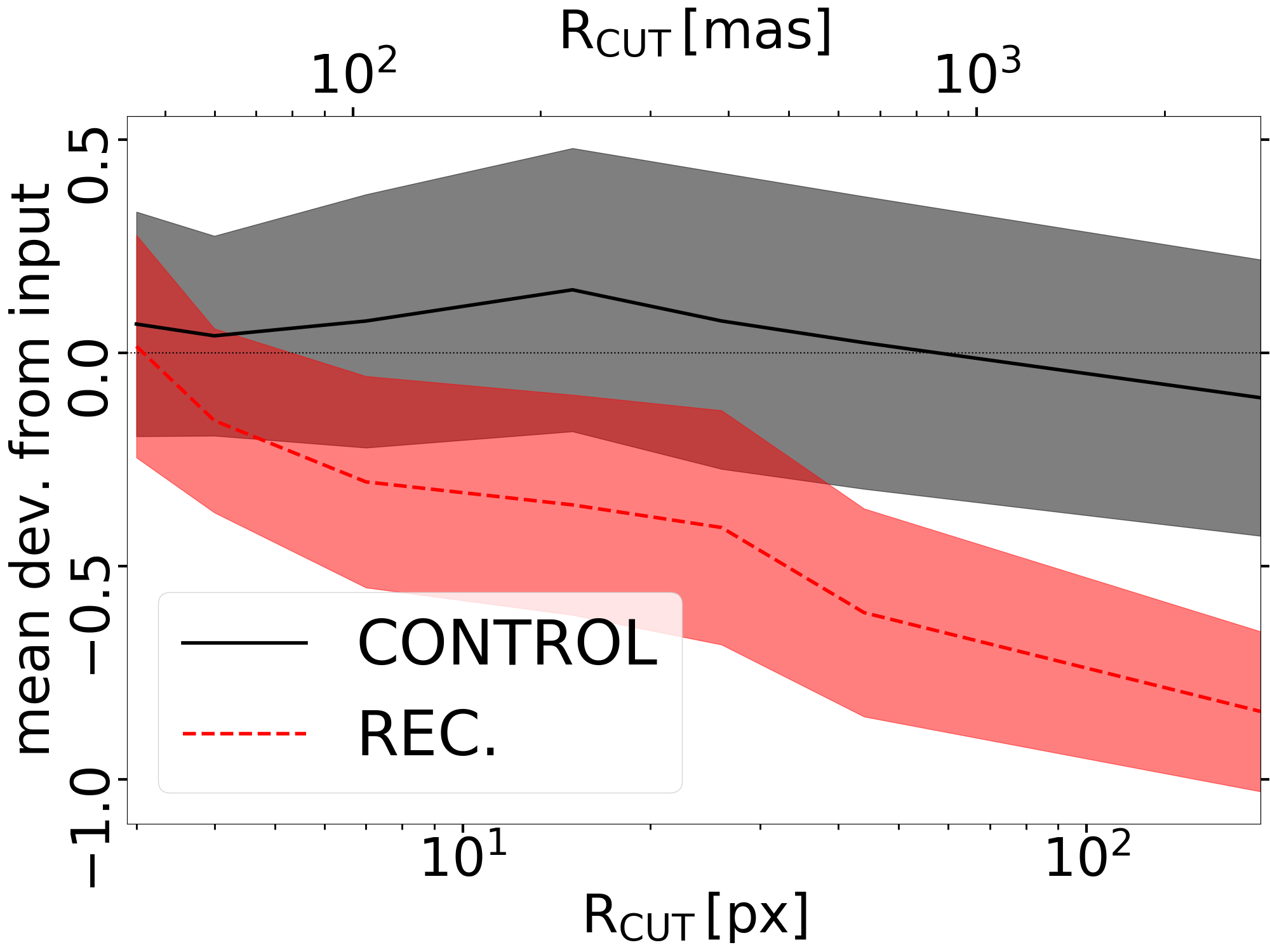} 
     \caption{Mean deviation from the input value of the measured S\'ersic index as a function of the cutting radius (${\rm R_{CUT}}$) used to mask the external regions of both oberved and reconstructed PSFs. While the PSFs cores are very similar, the small differences in the wings of the PSFs translate into a deviation of the measured S\'ersic index from the input value. After an initial settling to $-0.3$, the deviation resumes a shallow increase for ${\rm R_{CUT}}\gtrsim375\,{\rm mas}$.}
     \label{fig:sinxrp}
\end{figure}

We find that the only morphological parameter affected is indeed the S\'ersic index. The behaviour of the measured ${\rm n}$ as a function of ${\rm R_{CUT}}$ is reported in Figure \ref{fig:sinxrp}. If the central $29\,{\rm px}$ of the PSFs (${\rm R_{CUT}}=3\,{\rm px}=44.85\,{\rm mas}$) are considered, the differences between the reconstructed and observed PSFs are negligible.
Increasing ${\rm R_{CUT}}$, the recovered S\'ersic index progressively shifts towards lower values, suggesting that the bias in the measure of this morphological parameter is due to the non-perfect reconstruction of the wings of the observed PSF. Specifically, for ${\rm R_{CUT}}>45\,{\rm mas}$, the difference between measured and input $n$ values steeply increases up to $-0.3$ at ${\rm R_{CUT}}\lesssim90\,{\rm mas}$.
The recovered $n$ value remains almost constant up to ${\rm R_{CUT}}\lesssim0.4\,{\rm arcsec}$, from where a shallow increase in the deviation from input values is resumed.


\section{Summary and Conclusions}\label{sec:fin}

PSF reconstruction is a fundamental tool to analyze the data from AO instrumentation. It will support state-of-the-art scientific analysis of imaging and spectroscopic data. It is worth stressing here the fact that having a number of bright and isolated point sources in the observed field allows the PSF characterization directly from the science frames. This, however, can be tricky, in the case of AO observations where the PSF changes both with distance from the AO reference and in time.

The PSF-R technique allows to obtain the PSF of an astronomical observation simply from synchronous WFS telemetry data, without any access to the focal plane data. This is an unavoidable solution when there is no bright and isolated star in the instrumental FoV, suitable for deriving the observed PSF.

We have attempted here to carry out a PSF reconstruction of SOUL+LUCI LBT observations of bright, on-axis point sources in order to test the PSF-R code we are developing. 
To the best of our knowledge, this is the first published work to apply the PSF-R technique to real AO data acquired with a pyramid WFS. We have also carried out the PSF-R of an artificial source (an illuminated fiber), with optical turbulence disturbance generated and corrected using the Adaptive Secondary of LBT: these results have been already discussed in a preliminary stage by Ref.~\citenum{2020SPIE11448E..37S}. The controlled atmospheric conditions allow to check that the tool is able to successfully recover the injected perturbations.

The PSF-R software we have used is able to reproduce the observed FWHM, EE, and SR with an accuracy of 2-4\%, which is comparable to the level of accuracy reached by hybrid or calibrated techniques (see e.g. PRIME\cite{2020MNRAS.494..775B,2020A&A...634L...5M}). Moreover, the reconstructed PSF provides a better fit to the observed data than the diffraction limited one. This level of agreement points to a successful approach of the adopted PSF-R algorithm, which matches a technology readiness level equal to 7, following the international ranking defined by the ISO 16290/2013 standard.

With the present work, we aim at setting the foundations for forthcoming analysis and refinements of the adopted PSF-R method.
An effort is needed in order to improve the representation of the wings (at $>0.3\,{\rm arcsec}=20\,{\rm px}$) of the reconstructed PSF.
It is worth mentioning that even when bright stars are present in the FoV, empirical PSF methods would not be able to provide an accurate description of the PSF wings. In the near future, we plan focussing on the treatment of off-axis (anisoplanatism) corrections and testing the PSF-R algorithm in the case of crowded stellar fields, e.g. in the case of a globular cluster.

This kind of studies paves the way for exploiting AO data reduction for ELT-size telescope observations of dense stellar fields or individual extra-galactic objects. The PSF-R software shown here will be an important starting point for developing a similar software for MICADO, the first-light camera of ELT, which will adopt a pyramid WFS similar to SOUL in its SCAO observing mode. In light of the future implementation of the MAORY module on MICADO, our group is also contributing in extending the algorithm to MCAO systems. It is worth noting that in the ELT the halo of a star will be a factor of $\sim$5 larger than in LBT, and thus it will have a much lower surface brightness (a factor of $\sim 25$ fainter). On the other hand, the core of the PSF is a factor $\sim$5 narrower in diameter. This means that, especially at longer wavelengths, even for bright stars the halo will be barely detected as a distinct feature, and that multiple halos will be blended into a faint distributed background. This indicates that the PSF-R approach will be even more important in the future for 30-40m class telescopes. Only for stars that are bright enough to saturate, the outer halo becomes visible, thus the hybrid or calibrated PSF-R techniques will not be suitable in this case.

A first stage of scientific evaluation of the reconstructed PSFs has been carried out by simulating a long SOUL+LUCI LBT observation of a compact early type galaxy (red nugget) at $z\sim 1.5$. The reconstructed PSF allows us to derive the physical parameters (position, magnitude, half light radius) of the galaxy within 1 $\sigma$ level with GALFIT, while the S\'ersic index is reproduced with an accuracy of $\sim2\, \sigma$.
The axis ratio $b/a$ and position angle are reproduced within 5\% and 0.6 degree, respectively. We propose that the possible systematic shift of the S\'ersic index could be due to a non perfect representation of the wings of the reconstructed PSF with respect to the observed one. In summary, PSF reconstruction from telemetry data only is strongly required if one needs to obtain precise morphological measurements, since typical high-z galaxies are usually observed in blank extra-galactic fields. In this respect, the results of the presented analysis are extremely encouraging.

Lastly, we discuss, in the appendix, an interesting relation linking EE, the FWHM, and SR of the observed PSF that can be useful e.g. when one is interested in computing an instrument sensitivity.

We can conclude that the PSF-R approach presented here is an innovative method for pyramid WFS. 

\appendix
\section{A practical relation linking SR, FWHM, and EE}\label{sec:sr_vcorr}

In this appendix we present a useful relation that links the observed EE ($EE_{OBS}$) to the measured SR. Its application is of particular interest i.e. in the case that, for an AO systems, one is interested in the energy concentration in the core of the observed PSF knowing only the associated SR and FWHM with the final goal of computing an instrument sensitivity. 
The SR is defined as the ratio of the punctual peak values of the observed PSF and the diffraction limited one. A typical method for sensitivity computation is to consider the SR for the calculation of the Encircled Energy within the PSF core:
\begin{equation}\label{eq:wrongee}
    {\rm EE_{core,OBS,uncorrected}} \simeq {\rm SR} \cdot {\rm EE_{core,DL}} \,.
\end{equation}
In case of AO systems performing very well, e.g. SR$>30-50\%$, the core of the PSF can be assumed as the area within the first minimum of the  associated Airy function (i.e. the diffraction limited PSF).
Nevertheless, in these conditions, wrongly assuming that within this area one can find a SR-percentage of the flux corresponding to the Airy figure, brings to an underestimation of the flux within the aperture.

To overcome this problem, a correction can be made in the above equation by means of a factor that accounts for the actual PSF shape. On a first approximation, this shape factor corresponds to the combined effect of the very-low order modes corrected by the AO. The shape of the core is dominated by the residuals of these few modes. In real cases, tip-tilt residuals represent the majority of the uncorrected low-order fraction for two main reasons: i) because Kolmogorov power spectrum puts there $90\%$ of the energy and ii) because the largest non-turbulence wavefront error derives from telescope and instrument vibration (that may be as large as the atmospheric jitter and much faster).
Moreover, the correlation time of the atmospheric tip tilt - proportional to the ratio of wind velocity and telescope diameter - puts it in the $1-10\,{\rm Hz}$\cite{1992JOSAA...9..298C} , while the typical vibration power spectrum has peaks in the $10-100\,{\rm Hz}$ region.
When the PSF peak is similar to the diffraction limited regime, corresponding to relatively high SR (${\rm SR}>30\%-50\%$), one may correct the equation above by adding the shape-factor:
\begin{equation}\label{eq:eecorr}
\begin{aligned}
    {\rm EE_{core,OBS,corrected}} \simeq {\rm SR} \cdot {\rm EE_{core,DL}}\cdot \left(\frac{FWHM_{OBS}}{FWHM_{DL}}\right)^{2}
\end{aligned}
\end{equation}
and if opto-mechanical vibrations dominate the low-order residuals it can be restricted to the tip-tilt component
\begin{equation}\label{eq:eecorr2}
\begin{aligned}
    {\rm EE_{core,OBS,corrected}} \simeq {\rm SR} \cdot {\rm EE_{core,DL}}\cdot \left(\frac{FWHM_{VIBRATION}^{2}}{FWHM_{DL}^{2}}+1\right) \,,
\end{aligned}
\end{equation}
where the $FWHM_{VIBRATION}$ is the standard deviation of the jitter due to the vibration component.
In fact, under the hypothesis that both the residual vibrations component, uncorrected or introduced by the AO system, and the PSF core can be approximated by Gaussian functions, the broadening of the registered PSF is a function of the standard deviation associated to the vibration component. In particular, it is straightforward to derive that the peak value of the registered PSF is proportional to $FWHM_{OBS}^{-2}$.
It is also important to note that the ratio of EEs operates as a normalization factor, correlating with the amount of uncorrected photons outside the control radius of the AO system, defined as the distance from PSF centre where light distorted by uncorrected high-order spatial frequencies resumes following a seeing-limited distribution\cite{2003ApJ...596..702P} .

If applied to the nighttime dataset, Equation~\eqref{eq:wrongee} gives $EE_{core,OBS,uncorrected}=0.36$, in contrast with the observed value of $EE_{core,OBS}=0.43$. The shape factor in this case results:
\begin{equation}
\left(\frac{FWHM_{OBS}}{FWHM_{DL}}\right)^2=
\left(\frac{43.80}{40.07}\right)^2=1.19
\end{equation}

Thus, applying Equation~\eqref{eq:eecorr} it results $EE_{core,OBS,corrected}=0.43$ in perfect agreement with the observed value. From Equation~\eqref{eq:eecorr2} one can also compute the expected standard deviation associated to the residual vibration component. In the case of nighttime dataset, this amount to $7.4\,{\rm mas}$. For comparison, it results from telemetry that the amount of tip-tilt only in the WF residuals sum up to $3.9\,{\rm mas}$, indicating that other peak-broadening are likely occurring (i.e. defocus-astigmatisms, chromatic dispersion). 

\begin{figure}
    \centering
    \includegraphics[width=.7\columnwidth]{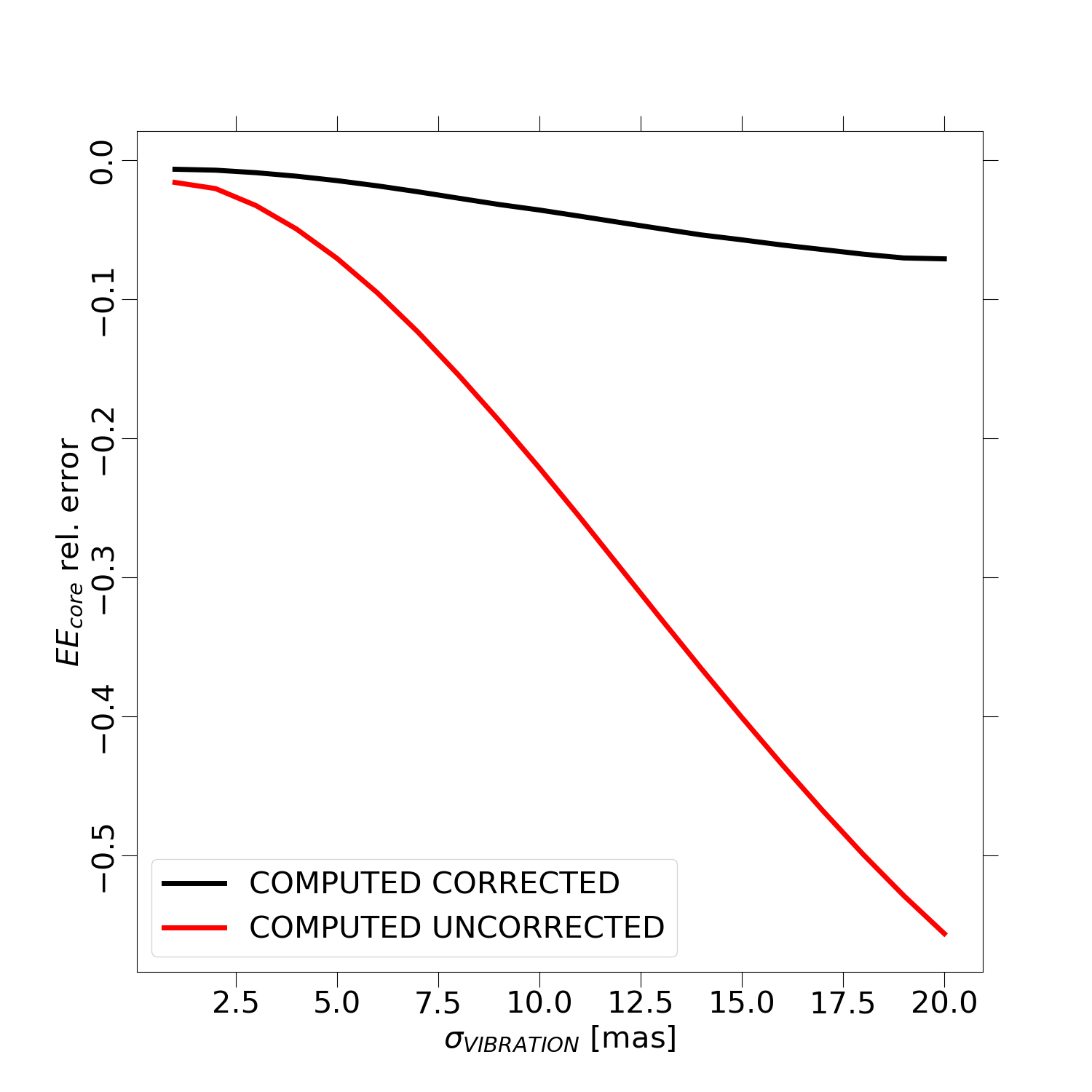}
    \caption{Application of Equation~\eqref{eq:wrongee} (UNCORRECTED) and Equation~\eqref{eq:eecorr} (CORRECTED) to mock diffraction limited PSFs, each convolved with a 2D Gaussian function of varying standard deviations. The relative error in the resulting EE are shown as a function of the standard deviation of the convolving 2D Gaussian ($\sigma_{VIBRATION}$).}
    \label{fig:eeatfw}
\end{figure}

Equation \eqref{eq:eecorr} provides accurate results only under particular conditions.
The major requirement is that the low-order aberration effect needs to be smaller than the EE aperture. Indeed, the higher is the broadening effect, the lower is EE inside a fixed aperture, therefore invalidating the normalization of the PSF. Specifically, the shape of the post-AO PSF inside the control radius matches as close as possible that of the diffraction limited PSF, but with intensity hampered by the amount of light outside the control radius associated to uncorrected modes. 

We propose that a simple solution to avoid this issue is to take care in defining the aperture radius for $EE_{core}$ of the order of $FWHM_{OBS}$. In other words, we encourage the use of aperture referring to actual, post-AO light distribution that implicitly takes into account the broadening due to residual vibrations. 

To quantify the limits of applicability of the relation in Equation~\eqref{eq:eecorr}, we performed a simple simulation. We convolved the diffraction limited PSF, oversampled by a factor $5$, with a 2D Gaussian functions of varying standard deviations (from $1$ to $20\,{\rm mas}$). In this case, as the observed PSF is obtained directly from the diffraction limited one, the lowering of SR values is only due to the residual vibration effects, which do not affect sensibly the wings of the PSF. Therefore, the effect of the aperture choice can also be estimated.
We then measured the associated $EE_{core,OBS}$ and $EE_{core,DL}$ inside an aperture of radius $R_{EE,core}=FWHM_{OBS}$. We also measured the SR directly from the resulting PSF. 
We then compared the measure $EE_{core,OBS}$ with those computed using Equations \eqref{eq:wrongee} and \eqref{eq:eecorr}. The results are reported in Figure~\ref{fig:eeatfw}. 
It can be observed that, after properly defining the aperture, Equation~\eqref{eq:eecorr} correctly estimates the observed $EE_{core,OBS}$ within $10\%$ accuracy for a wide range of residual vibrations.


\subsection*{Acknowledgements}
We thank the anonymous referees for their thorough revision that enhanced the quality of the present work. This work has been partly supported by INAF through the Math, ASTronomy and Research (MAST\&R), a working group for mathematical methods for high-resolution imaging. 
Based on observations made at the Large Binocular Telescope
(LBT) at Mt. Graham (Arizona, USA).
The LBT is an international collaboration among institutions in the United States, Italy and Germany. LBT Corporation partners are: The University of Arizona on behalf of the Arizona university system; Istituto Nazionale di Astrofisica, Italy; LBT Beteiligungsgesellschaft, Germany, representing the Max-Planck Society, the Astrophysical Institute Potsdam, and Heidelberg University; The Ohio State University, and The Research Corporation, on behalf of The University of Notre Dame, University of Minnesota and University of Virginia.


\bibliography{lbtpsfr.bib}
\bibliographystyle{spiejour}   

\end{document}